\documentclass[onecolumn,authoryear]{els-mrw} 

\newcommand{\msun}{\,\mathrm{M}_\odot}
\newcommand{\rsun}{\,\mathrm{R}_\odot}
\newcommand{\Myr}{\,\mathrm{Myr}}
\newcommand{\au}{\,\mathrm{AU}}
\newcommand{\cm}{\,\mathrm{cm}}
\newcommand{\pcmc}{\,\mathrm{cm}^{-3}}
\newcommand{\pc}{\,\mathrm{pc}}
\newcommand{\K}{\,\mathrm{K}}
\newcommand{\kms}{\,\mathrm{km\,s^{-1}}}
\newcommand{\kmps}{\,\mathrm{km\,s^{-1}}}

\newcommand{\sigv}{\sigma_v}
\newcommand{\cs}{c_\mathrm{s}}
\newcommand{\mach}{\mathcal{M}}
\newcommand{\sigs}{\sigma_s}
\newcommand{\means}{\langle s\rangle}
\newcommand{\meanrho}{\langle\rho\rangle}
\newcommand{\sfr}{\mathrm{SFR}}
\newcommand{\sfrff}{\mathrm{SFR_{ff}}}
\newcommand{\tff}{t_\mathrm{ff}}
\newcommand{\mcl}{M_\mathrm{cloud}}
\newcommand{\scrit}{s_\mathrm{crit}}
\newcommand{\eps}{\epsilon}
\newcommand{\phit}{\phi_t}
\newcommand{\phix}{\phi_x}
\newcommand{\alphavir}{\alpha_\mathrm{vir}}

\usepackage[tight,k-tight]{minitoc}

\usepackage{amsmath,amssymb,amsfonts,amsthm,makeidx,graphicx}
\usepackage{txfonts}
\usepackage{helvet}
\usepackage{cleveref}
\usepackage{subcaption,graphicx}
\mtcsetfeature{minitoc}{open}{\vspace{-0.3cm}}
\usepackage{tcolorbox}

\usepackage{ulem}
\usepackage{xcolor}

\begin{document}

\dominitoc
\chapter{Star Formation}
\label{chap:star_formation}

\author[1]{Rajika Kuruwita}%
\author[2]{\L ukasz Tychoniec}%
\author[3]{Christoph Federrath}%

\address[1]{\orgname{Heidelberg Institute for Theoretical Studies}, \orgdiv{Stellar Evolution Theory}, \orgaddress{Schlo{\ss}-Wolfsbrunnenweg 35, 69118 Heidelberg, Germany}}
\address[2]{\orgname{Leiden University}, \orgdiv{Leiden Observatory}, \orgaddress{PO Box 9513, 2300RA, Leiden, The Netherlands}}
\address[3]{\orgname{Australian National University}, \orgdiv{Research School of Astronomy and Astrophysics}, \orgaddress{ACT 2611, Australia}}

\articletag{Chapter Article tagline: update of previous edition, reprint..}

\maketitle

\mtcsetfeature{minitoc}{after}{\vspace{-20pt}}
\mtcsetfeature{minitoc}{before}{\vspace{-20pt}}
\minitoc

\begin{glossary}[Glossary]
\term{Molecular cloud} a region of space primarily composed of hydrogen in its molecular state (H$_2$).\\
\term{Pre-stellar core} an over-density of gas that does not have a star, but will collapse to form a star.

\end{glossary}

\begin{glossary}[Nomenclature]
\begin{tabular}{@{}lp{34pc}@{}}
MHD & Magnetohydrodynamics\\
ISM & Interstellar medium\\
AU & Astronomical units\\
YSO & Young stellar object\\
PDF & Probability density function\\
SED & Spectral energy distribution\\

\end{tabular}
\end{glossary}

\begin{abstract}[Abstract]
In this chapter, we will cover how stars form from the stellar nurseries that are giant molecular clouds. We will first review the physical processes that compete to regulate star formation. We then review star formation in turbulent, magnetized molecular clouds and the associated statistics giving rise to the star formation rate and the initial mass function of stars. We then present the protostellar stages in detail from an observational perspective. We will primarily discuss low-mass ($<1.5\msun$) stars. Finally, we examine how multiplicity complicates the single-star formation picture. This chapter will focus on star formation at redshift~0.

\textbf{Key words}: Astrophysical processes: astrophysical magnetism – gravitation; Interstellar medium: interstellar dynamics, molecular clouds; Protostars: young stellar objects.
\end{abstract}

\begin{tcolorbox}[
 standard jigsaw,
 colback=green!15,
 opacityframe=0.5,
 opacityback=0.2
]
\section*{Learning Objectives}
\begin{itemize}
  \item Understand the role of gravity, hydrodynamics, turbulence, radiation, and magnetic fields in star formation.
  \item Look at how the star-forming environments set the initial conditions for star formation and the rate of star formation.
  \item Review what observations of young stellar objects (YSOs) tell us about the star formation process.
  \item Explore how the picture of isolated star formation is complicated by the fact that most stars are born with siblings.
\end{itemize}

\end{tcolorbox}

\section{Introduction to the physics of star formation}
\label{sec:physical_processes}

The most abundant element in the universe is hydrogen (see `Big Bang Nucleosynthesis'), which typically exists in an ionized state as protons in low-density, hot regions (temperature $T>10^4\K$), or as atomic hydrogen with a proton and electron in higher-density, cooler ($100\K\lesssim T<10^4\K$) regions \citep{Ferriere2001,McClureGriffithsEtAl2023}. However, when the density is high enough and the temperature is sufficiently low \citep[$\sim10-100\,\mathrm{K}$; see e.g.,][]{tacconi_evolution_2020}, hydrogen atoms can form the more stable H$_2$ molecule. These regions of higher density, where molecular hydrogen exists, are called `molecular clouds'. Stars form from over-densities of gas within these molecular clouds.

\subsection{Gravity, gas dynamics and the formation of the first hydrostatic core}
\label{ssec:gravity}

When a local over-density forms in a molecular cloud, it is typically called a `pre-stellar core'. At this stage, a star has not formed, but the pre-stellar core will collapse under its own gravity to start the star-formation process. Assuming that the pre-stellar core is mostly spherical, with no rotation, gravity is the only force acting on the pre-stellar core. Also assuming that the gas is pressure-less, a star should form within a free-fall time $\tff$. The free-fall time for a spherical cloud of uniform density $\rho$ is defined as

\begin{align}
\label{eq:t_ff}
\tff =\sqrt{\frac{3\pi}{32G\rho}},
\end{align}

\noindent where $G$ is the gravitational constant. Observed pre-stellar cores have number densities of $n_{\mathrm{H2}} \simeq 10^5\cm^{-3}$ \citep{keto_different_2008}, which translates to $\rho \simeq 3.8 \times 10^{-19}\,\mathrm{g}\cm^{-3}$. At this density, the free-fall time of a typical pre-stellar core is $0.1\Myr$.

However, gas is not a pressure-less fluid, and as the pre-stellar core collapses, the outward force of the pressure gradient will counteract the inward gravitational force. When the collapse progresses to gas densities of $\simeq10^{-10}\,\mathrm{g}\cm^{-3}$, the pressure force and the gravitational force are comparable and the fluid is approaching hydrostatic equilibrium. This is defined as

\begin{align}
\label{eq:hydrostatic}
\frac{dP(r)}{dr} = \frac{GM(r)\rho(r)}{r^2},
\end{align}

\noindent where $P(r), \rho(r)$ and $M(r)$ are the pressure, density, and enclosed mass, at radius $r$. This hydrostatic object that forms from the pre-stellar collapse is called the `first hydrostatic core' \citep{larson_numerical_1969}. This first hydrostatic core is not yet a star, because its radius is $\sim$4$\au$. To describe the further collapse of the first hydrostatic core into a star, we must now understand how radiation plays a role.

\subsection{Radiation and the formation of the second hydrostatic core}
\label{ssec:radiation}

Electromagnetic radiation can be produced from two pathways: 1.~Converting mass into radiative energy through stellar nucleosynthesis/nuclear fusion (using Einsteins $E=mc^2$; also see `Evolution and final fates of low- and intermediate-mass stars', `Evolution and final fates of massive stars'), or 2.~Converting from a different type of energy into radiative energy. 

Conversion between energy types is an essential physical process. During the pre-stellar core collapse described in \Cref{ssec:gravity}, the gas has gravitational potential energy that is converted into thermal energy. This thermal energy is equivalent to the gas pressure, which eventually slows down the pre-stellar collapse, to form the first hydrostatic core. The question then is, how do we continue to collapse beyond the first hydrostatic core to form a star?

Thermal energy can be removed from the system via radiation. During the initial stages of pre-stellar collapse, the thermal energy from the gravitational collapse can be radiated away efficiently because the gas has a low optical depth ($\tau$). This efficient radiative cooling means that the gas remains approximately isothermal until number densities of $\sim10^{10}\pcmc$ are reached.

The optical depth is a measure of how easily photons can pass through a medium, telling us if a medium is opaque (or `optically thick'; $\tau\gg1$) or transparent (or `optically thin'; $\tau\ll1$). The optical depth is defined as

\begin{align}
\label{eq:optical_depth}
\tau_\lambda = \int_{0}^{L} n\sigma_\lambda dl,
\end{align}

\noindent where $\sigma_\lambda$ is the cross-section of interaction of the particles in the medium with photons of wavelength $\lambda$, and $n$ is the number density of particles. The optical depth is calculated by integrating these values over some line of sight $L$. 

The interactions that are quantified by $\sigma_\lambda$ are absorption and scattering events. An atom or molecule is more likely to absorb a photon if its energy matches a difference in electron shell energies because this can excite electrons into higher energy states. This process produces absorption-line spectra. Scattering events are dependent on the size of the particles ($r_p$) and wavelength. In the Rayleigh scattering regime, longer wavelength photons can penetrate the medium further before an interaction than shorter wavelength photons. For a hydrogen molecule, which has a size of $1.2\times10^{-8}\cm$, any wavelength from the visible light ($\sim$5$\times10^{-5}\cm$) to radio waves ($>10^2\cm$) falls into the Rayleigh scattering regime.

In the optically thin regime, a photon of radiation can pass through gas unhindered allowing for efficient radiative cooling. As stated previously, the temperature of molecular hydrogen in molecular clouds is quite cool, around $10-40\,\mathrm{K}$, which according to Wien's law would radiate a blackbody spectrum that peaks in the infrared ($\sim$10$^{-2}\cm$). During the initial protostellar collapse, the number density of molecular hydrogen is low enough that the infrared radiation is optically thin, leading to efficient cooling.

In the optically thick regime, a photon has many interactions with particles in the medium being absorbed, re-emitted, and scattered. The trapping of radiation makes cooling very inefficient, and radiation is more likely to be converted to thermal energy increasing the temperature of the medium. This is what happens when the first hydrostatic core forms: the medium transitions from optically thin to optically thick.

The first hydrostatic core is still accreting mass from the surrounding pre-stellar core, which disturbs the hydrostatic equilibrium. With the increasing mass, the hydrostatic core will continue to contract under its own gravity. However, because the gas is optically thick, making radiative cooling inefficient, the temperature of the core increases during contraction.

Once the temperature at the center of the first hydrostatic cores exceeds $2000\,\mathrm{K}$, the H$_2$ molecules separate into individual atoms. When this dissociation happens, thermal energy is used to break the chemical bonds in the H$_2$ molecular to create atomic hydrogen. Because of this efficient use of thermal energy for H$_2$ dissociation, the system returns to a near isothermal state and a second isothermal collapse is triggered. As with the first isothermal collapse, the second isothermal collapse will halt when the internal pressure increases and a new hydrostatic equilibrium is established. 

The second hydrostatic core is a few solar radii large and is embedded in the first hydrostatic core, which the second core accretes from. Further accretion makes the second hydrostatic core undergo another adiabatic contraction, which increases the central density and temperature, ionizing the atomic hydrogen. When the central temperature exceeds $10^4\,\mathrm{K}$ the hydrogen is fully ionized. The higher density and temperature make the second core opaque to the radiation being produced in the center. Convection is triggered within the second hydrostatic core because of the high optical depth.

When most of the mass from the pre-stellar core has been accreted by the second core, we essentially have a protostar. This protostar will continue to contract, following the Hayashi track, decreasing in luminosity, but maintaining the same surface temperature of $\sim$4000$\,\mathrm{K}$ \citep{hayashi_evolution_1966}. For low-mass protostars ($M_\star<0.6\msun$), they remain fully convective and will continue contracting along the Hayashi track until hydrogen fusion is triggered, and the star is on the main sequence. For more massive protostars, the central temperature of the protostar increases more during the Hayashi track contraction, and the core becomes radiative. These stars continue to contract slowly, but the surface temperature increases, effectively making the protostar have a constant luminosity \citep{henyey_early_1988}, eventually also triggering hydrogen fusion and joining the main sequence. For a more detailed evolution of the later stages of star formation. For further details on the stellar evolution of stars, we refer the reader to `Evolution and final fates of low- and intermediate-mass stars' and `Evolution and final fates of massive stars'.

\subsection{Magnetic fields and angular momentum transport}
\label{ssec:magnetic_fields}

In the previous section, we focused on the collapse of a non-rotating cloud. However, the pre-stellar cores that form in giant molecular clouds inherit angular momentum from the turbulence in the parent cloud, such that the cores are always rotating. Even though these clouds initially have very low rotation, as they collapse, they rotationally flatten and spin up due to the conservation of angular momentum \citep{tscharnuter_collapse_1987}. The collapse halts when the radius of the cloud matches the centrifugal radius. This is defined as the radius where the rotation rate equals the Keplerian velocity and is given by

\begin{align}
\label{eq:centrifugal_radius}
r_c = \frac{j^2}{GM},
\end{align}

\noindent where $M$ and $j$ are the mass and mass-specific angular momentum of the pre-stellar core, respectively. However, based on observed rotation rates, for a $1\msun$ pre-stellar core, the centrifugal radius is $2.7\times10^6\rsun$, which is significantly larger than a star, i.e. $\sim$1$\rsun$ \citep{spitzer_physical_1978, tomisaka_evolution_2000}. Therefore, approximately 99.99\% of the angular momentum must be removed, for the gas to collapse down to stellar radii. This angular is now understood to be removed via jets and outflows which are the result of magnetic fields.

Magnetic fields permeate the universe and play an important role in the star formation process. Charged particles preferentially travel along magnetic fields. As stated earlier, most of the gas in the ISM exists in an ionized state, which is essentially full of positively charged protons, and negatively charged electrons. Due to the charged nature of the ISM, it is expected that there is a strong coupling between gas motions and magnetic fields. However, even in regions where the ionization fraction is very low, such as in the star-forming, molecular phase of the ISM, the coupling between ions and neutrals ensures that even the neutral gas is subject to the same Lorentz force as the ions.

Magnetic field lines have tension, similar to a rubber band. If a magnetic field line is perturbed by charged gas moving perpendicular to the line, the magnetic tension makes the field line want to return to its unperturbed state. The magnetic field line, bouncing back after being perturbed, excites waves that move along the magnetic field. These waves are called Alfv\'en waves, and they can remove angular momentum and energy from a collapsing pre-stellar core. As pre-stellar cores rotate and collapse, the gas drags the magnetic field along with the motion, building magnetic pressure. Magnetic pressure is defined as

\begin{align}
\label{eq:magnetic_pressure}
P_\mathrm{B} = \frac{B^2}{2\mu_0}.
\end{align}

\noindent where $B$ is the magnetic field strength, $\mu_0$ is the vacuum permeability. The force from the magnetic pressure gradient is the actual restoring force for the magnetic field lines and acts as the `tension' mentioned previously. This pressure gradient force $F_\mathrm{B}$ opposes the gravitational force, and this pressure gradient causes the field lines to `snap' back triggering Alfv\'en waves to travel along magnetic field lines with velocity $v_A$, and reduces the rotational speed of the pre-stellar core. The velocity of Alfv\'en waves is given by

\begin{align}
\label{eq:alfven}
v_A = \frac{B}{\sqrt{\mu_0\rho}},
\end{align}

\noindent $\rho$ is the mass density of the gas.

The energy for the Alfv\'en waves is converted from the kinetic energy of the ionized material in the pre-stellar core, and this is how magnetism can remove energy and angular momentum during the initial stages of pre-stellar core collapse, however at later stages other magnetic mechanisms become more dominant.

While magnetism can remove energy from moving ionizing gas in a collapsing pre-stellar core, strong magnetic fields can slow down, or halt collapse due to magnetic pressure. As described in \Cref{ssec:gravity}, the collapse of the pre-stellar core due to gravity is counteracted by the gas pressure. Magnetic pressure also contributes to counteracting gravitational collapse. The mass-to-flux ratio ($M/\Phi$) has historically been used to quantify whether a pre-stellar core will collapse, or if the magnetic pressure is too high that it prevents collapse, where $\Phi$ is the magnetic flux through the pre-stellar core. This ratio essentially quantifies the balance of the gravitational force against the force due to magnetic pressure. The critical mass-to-flux ratio ($(M/\Phi)_\mathrm{crit}$) was calculated by \citet{mouschovias_note_1976} to be $487\,\mathrm{g}\,\mathrm{cm}^{-2}\,\mathrm{G}^{-1}$, where cores that are supercritical (i.e. $(M/\Phi)>(M/\Phi)_\mathrm{crit}$) will collapse, while sub-critical cores ($(M/\Phi)<(M/\Phi)_\mathrm{crit}$) will not.

\begin{figure}[b]
\centering
\includegraphics[width=\textwidth]{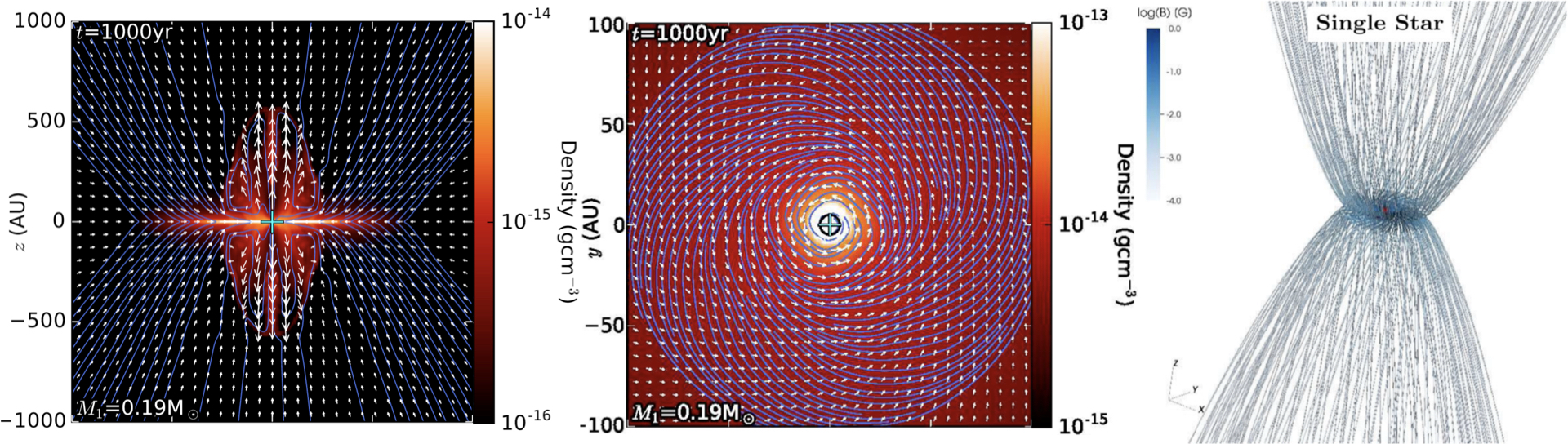}
\caption{Magnetic field morphology, side-on (left) and top-down (middle) in a simulation of protostar formation in an accretion disc. A three-dimensional rendering of this morphology is shown in the right panel. Adapted from figures 2, 5, and 11 of \citet{kuruwita_binary_2017}.}
\label{fig:b_fields}
\end{figure}

\begin{figure}[t]
\centering
\includegraphics[width=0.6\textwidth]{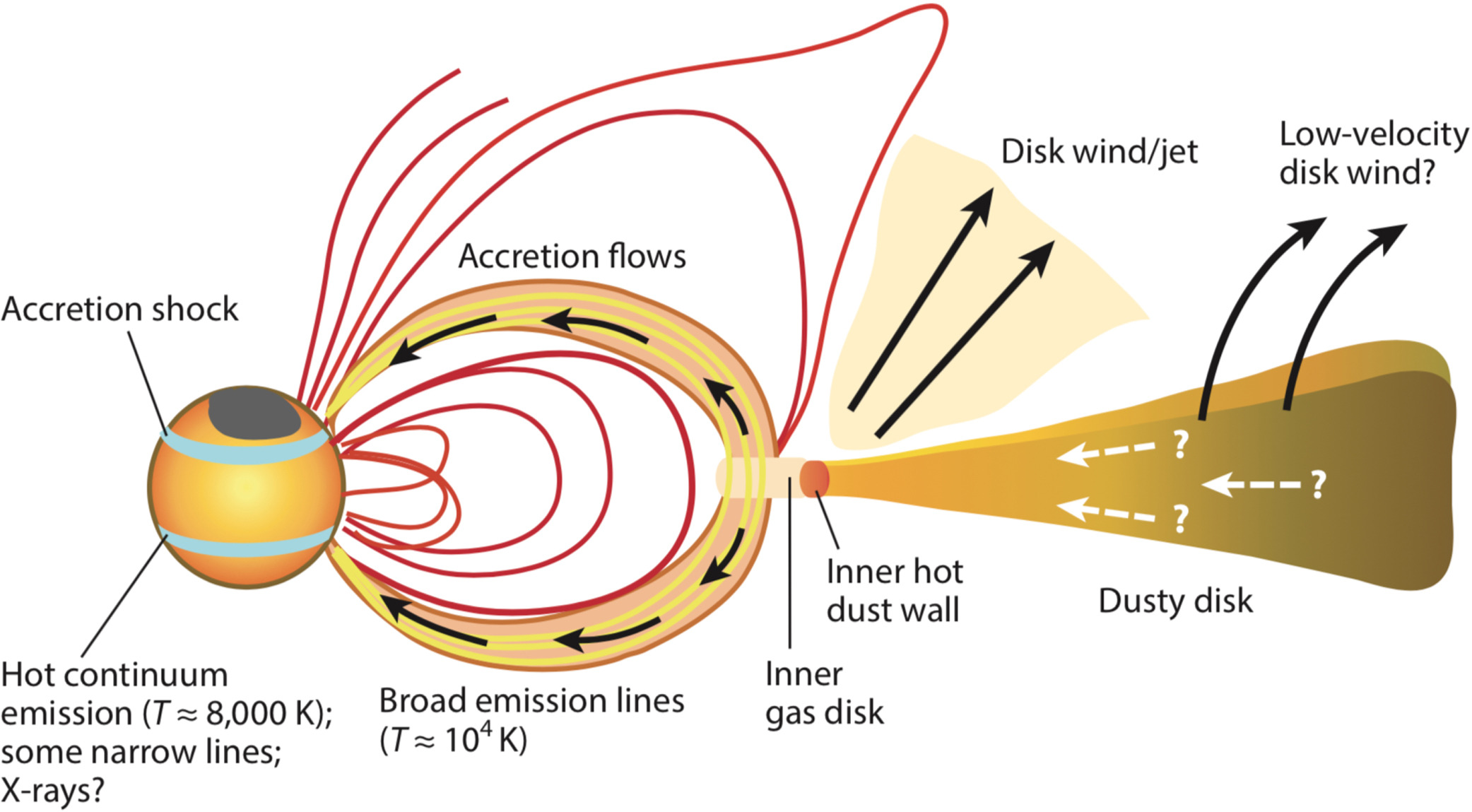}
\caption{Schematic view of a young star accreting from a disk through the stellar magnetosphere. Jets are launched from the inner disk, while disk winds are launched at larger radii. Both mechanisms remove angular momentum, allowing the gas to move inwards through the disk. The protostellar magnetic field threads through the inner disk, allowing ionized material to be funneled along these lines onto the protostar. Figure 1 from \citet{hartmann_accretion_2016}, used with permission.}
\label{fig:disk_winds}
\end{figure}

As supercritical pre-stellar cores collapse, the rotation leads to the formation of a disk due to the conservation of angular momentum, but the magnetic field is also dragged along with the gas and is wound up within the disk. \Cref{fig:b_fields} shows this magnetic field morphology produced from simulations of the collapse of a rotating supercritical core. The left and middle panels show gas density projections of this core perpendicular and parallel to the rotation axis of the core, and the blue streamlines trace the magnetic field. The right panel shows a three-dimensional rendering of the magnetic field, with the color indicating magnetic field strength. We see that the magnetic field morphology is pinched inwards in the disk, and within the disk, the magnetic field is coiled up. A natural consequence of the coiling of magnetic fields is the production of protostellar jets and outflows. Multiple mechanisms have been proposed for launching outflows including the disk wind model \citep{blandford_hydromagnetic_1982, konigl_disk_2000}, the magnetic tower \citep{lynden-bell_why_2003} and the `X-wind' model \citep{shu_magnetocentrifugally_1994}. All of these models are likely to be present during the star formation process and act in different regimes. \Cref{fig:disk_winds} describes how the protostar and circumstellar disk are connected, and where different outflows are launched.

The \emph{disk wind model}: \citet{blandford_hydromagnetic_1982} calculated that centrifugally-driven outflows can be launched from a disk with a coiled-up magnetic field if the angle of the magnetic field to the disk mid-plane is less than $60^\circ$. The velocity of the outflows reflects the rotation profile of the disk, with faster-velocity outflows being launched at smaller radii, and lower-velocity outflows being launched at larger radii. These outflows are often called `winds' and are launched from the disk surface, as shown in \Cref{fig:disk_winds}.

The \emph{magnetic tower} describes the launching of outflows via a magnetic pressure gradient. \citet{lynden-bell_why_2003} describe these as highly coiled-up magnetic structures and the pinching of magnetic fields to produce strong pressure gradients away from the disk, producing a force that significantly overcomes the gravitational force. Many ideal MHD simulations of molecular core collapse and protostar formation find that this mechanism is what drives the initial jet launching \citep{banerjee_outflows_2006, machida_impact_2012}. The magnetic tower is also likely acting along with the magneto-centrifugally driven disk winds of \citet{blandford_hydromagnetic_1982, nolan_centrifugally_2017}.

The \emph{X-wind model} mainly concerns regions in the inner disk where the magnetosphere of the star threads through the inner disk, as shown in \Cref{fig:disk_winds}. Because of the heat from the protostar, the inner disk is fully ionized, and strong coupling between the gas and magnetic field. Due to the conservation of angular momentum, the protostar would typically rotate quickly, but because of this strong coupling between the protostar and inner disk which rotates at lower speeds, (magnetic) tension builds up between these two regions. This tension leads to the launching of jets by the X-wind mechanism. These jets have velocities of a few $100\,\mathrm{kms}$, similar to velocities observed in protostellar jets.

The modern consensus is that it is a combination of the disk wind and X-wind model working together to produce the outflow features observed in protostars, with the X-wind producing the highly collimated jet and the disk-wind producing the lower-velocity outflow. Overall, magnetic fields play an important role in the removal of angular momentum, aiding pre-stellar collapse, and allowing protostars to accrete from their circumstellar disks.

\section{Stellar nurseries -- molecular clouds}
\label{sec:molecular_clouds}

Stars form in cold, turbulent, molecular clouds. We know this from molecular-line observations with radio and sub-mm telescopes (see also 'The interstellar medium'). These clouds consist mainly of molecular hydrogen, H$_2$, with carbon monoxide, CO, being the second-most abundant molecule. CO is typically used to measure the turbulent velocities in molecular clouds, because at the low temperatures of about \mbox{$10$--$50\K$}, H$_2$ cannot emit photons due to its missing permanent dipole moment, while CO is easily excited, and the Doppler shift of its rotational lines can be used to measure the line-of-sight (LOS) velocity of the gas \citep{StahlerPalla2004}.

Given the typical temperatures of molecular clouds, implying sound speeds of the order of \mbox{$\cs\approx0.2$--$0.5\kmps$}, and measured velocity dispersion of \mbox{$\sigv\approx0.5$--$10\,\kmps$}, the clouds are governed by supersonic turbulent motions with sonic Mach numbers of $\mach=\sigma_v/\cs\approx1$--$50$. When studied over different length scales, the velocity dispersion follows a power-law relation with scale, $\ell$,
\begin{equation} \label{eq:veldipsizerel}
\sigv(\ell) = \sigv(L)\,\left(\frac{\ell}{L}\right)^p \approx 1\kmps\,\left(\frac{\ell}{\mathrm{pc}}\right)^p,
\end{equation}
with $p\approx0.4$--$0.5$ based on observations \citep[e.g.,][]{Larson1981,SolomonEtAl1987,OssenkopfMacLow2002,HeyerBrunt2004,RosolowskyBlitz2005,HeyerEtAl2009,RomanDuvalEtAl2011}. This power-law form is indeed similar to the power law obtained in the famous Kolmogorov model of turbulence ($p=1/3$) \citep[][]{Kolmogorov1941c,Frisch1995}, however, which strictly only applies to incompressible turbulence. Instead, the Burgers model of turbulence \citep{Burgers1948} may be more applicable here, as it is based on an ensemble of discontinuities (shocks), which corresponds to $p=1/2$. Reality likely sits in between those extremes, not to mention the added complication of intermittency and magnetic fields, with active research exploring turbulence models for this complex regime of compressible plasma turbulence \citep{SheLeveque1994,BoldyrevNordlundPadoan2002,BrandenburgSubramanian2005,SchekochihinEtAl2007,SchmidtFederrathKlessen2008,KonstandinEtAl2012,Federrath2016jpp,SetaFederrath2021,AchikanathEtAl2021,FederrathEtAl2021,BeattieEtAl2023}.

\subsection{Turbulence-regulated star formation} 
\label{sec:sfrbasics}

Motivated by the fact that all star-forming clouds observed so far exhibit high levels of compressible turbulence, many authors have investigated star formation in turbulent media \citep{Padoan1995,KlessenHeitschMacLow2000,ElmegreenEtAl2003,KrumholzMcKee2005,PadoanNordlund2011,HennebelleChabrier2011,FederrathKlessen2012,Federrath2018,BurkhartMocz2019}. The compressible nature of turbulence gives rise to a characteristic gas density probability distribution function (PDF), enabling analytic estimates of the star formation rate (SFR) and the initial mass function (IMF; See also `Stellar initial mass function') of stars.

\subsubsection{The gas density PDF} 
\label{sec:pdf}

Turbulent isothermal gas can be well approximated by a log-normal density PDF \citep{Vazquez1994,PassotVazquez1998,PadoanNordlund2002,KritsukEtAl2007},
\begin{equation} 
\label{eq:pdf}
p(s) = \left(2\pi \sigs^2\right)^{-1/2}\exp\left[-\frac{\left(s-\means\right)^2}{2\sigs^2}\right],
\end{equation}
with the dimensionless logarithmic density contrast $s=\ln(\rho/\meanrho)$, mean density $\meanrho$, mean log-density $\means=-\sigs^2/2$ \citep{LiKlessenMacLow2003,FederrathKlessenSchmidt2008,FederrathDuvalKlessenSchmidtMacLow2010}, and log-density variance \citep{PadoanNordlund2011,MolinaEtAl2012},
\begin{equation} 
\label{eq:sigs}
\sigs^2=\ln\left[1+b^2\mach^2\left(1+\beta^{-1}\right)^{-1}\right],
\end{equation}
where $\beta$ is the plasma beta (ratio of thermal to magnetic pressure; note that $\beta\to\infty$ if the magnetic field is zero). A more recent modification of this relation accounts for strong magnetic guide fields \citep{BeattieEtAl2021}. The parameter $b$ in \Cref{eq:sigs} is the turbulence driving parameter, which is controlled by the mixture of solenoidal vs.~compressive modes in the driving mechanism of the turbulence \citep{FederrathKlessenSchmidt2008}. Purely solenoidal (divergence-free) driving has $b\sim1/3$, while purely compressive (curl-free) driving has $b\sim1$ \citep{FederrathDuvalKlessenSchmidtMacLow2010,DhawalikarEtAl2022,GerrardEtAl2023}. Modifications of \Cref{eq:sigs} can be made to account for non-isothermal gas conditions \citep{NolanFederrathSutherland2015,FederrathBanerjee2015}, and where intermittency plays a role \citep{Hopkins2013PDF}.

\subsubsection{The star formation rate}
\label{sec:sfr}

The modern theory of star formation is based on the turbulent density PDF. A key step in turbulence-regulated theories of the SFR and IMF is to estimate the fraction of dense gas that can form stars, and this is exactly what \Cref{eq:pdf} and~(\ref{eq:sigs}) can provide. To derive a rate at which this dense gas turns into stars, we need to divide the dense gas fraction by the freefall time, which, using the definitions in Sec.~\ref{sec:pdf}, gives the basic expression for the SFR per average freefall time, $\langle\tff\rangle$, for a cloud of mass $\mcl$ \citep[see][]{HennebelleChabrier2011,FederrathKlessen2012},
\begin{equation} 
\label{eq:sfrff}
\sfrff = \sfr\frac{\langle\tff\rangle}{\mcl} =
\frac{\eps}{\phit} \int_{\scrit}^{\infty} \frac{\rho}{\langle\rho\rangle} \frac{\langle\tff\rangle}{\tff(\rho)}\,p(s)\,ds =
\frac{\eps}{\phit} \int_{\scrit}^{\infty}\exp\left(\frac{3}{2}s\right)\,p(s)\,ds = 
\frac{\eps}{2\phit} \exp\left(\frac{3}{8}\sigs^2\right) \left[1+\mathrm{erf}\left(\frac{\sigs^2-\scrit}{(2\sigs^2)^{1/2}}\right)\right],
\end{equation}
where $\tff(\rho)=3\pi/(32G\rho)$ defined in \Cref{eq:t_ff}, the star-to-core mass ratio \mbox{$\eps\sim0.3$--$0.5$} \citep{MatznerMcKee2000,FederrathEtAl2014}, and $\phit\sim2$ is a numerical correction factor, calibrated in simulations \citep[see tab.~3 in][]{FederrathKlessen2012}. Finally, the critical density for star formation, $\scrit$, is given by 
\begin{equation} 
\label{eq:scrit}
\scrit = \ln\left[(\pi^2/5)\,\phix^2\,\alphavir\,\mach^{2}\left(1+\beta^{-1}\right)^{-1}\right],
\end{equation}
which is obtained by comparing the Jeans length \citep{KrumholzMcKee2005} with the turbulent sonic scale \citep{FederrathEtAl2021}, which marks the transition from supersonic turbulence on cloud scales, to subsonic turbulence inside the dense star-forming cores and accretion disks. Thus, the critical density is a result of the competition of gravity and turbulence, which gives rise to the virial parameter $\alphavir=2E_\mathrm{kin}/E_\mathrm{grav}$ in \Cref{eq:scrit}, the ratio of twice the kinetic to gravitational energy of the cloud. The numerical correction factor $\phix\sim0.2$ accounts for a slight mismatch between the sonic and Jeans scales when forming $\scrit$, and can be determined by calibration with numerical simulations \citep[see tab.~3 in][]{FederrathKlessen2012}.

A key prediction of this theory is that the SFR depends on 4~basic cloud parameters, namely the virial parameter ($\alphavir$), the sonic Mach number ($\mach$), the turbulence driving mode ($b$), and the magnetic plasma beta ($\beta$). For instance, keeping all parameters fixed at typical cloud values ($\alphavir\sim1$, $\mach\sim10$, $\beta\sim0.3$), except for the driving mode, \Cref{eq:sfrff} predicts an SFR that is a factor of $\sim2.4$ higher for compressive driving compared to solenoidal driving. With the associated reduction in $\alphavir$ for compressive driving, due to the stronger local compressions leading to a higher overall binding energy of a cloud, compressive driving can yield an order of magnitude higher SFR than solenoidal driving \citep{FederrathKlessen2012}.

\subsection{The initial mass function of stars}
\label{ssec:IMF}

The initial mass function (IMF) is the distribution of the birth mass of stars. `Stellar initial mass function' in this series provides more information, so we focus here on the link between turbulence and the IMF and present a brief summary of the physics primarily involved in controlling the IMF.

\subsubsection{Basic characterization}

The IMF is usually characterized by a power-law section for masses $\gtrsim1\msun$ \citep{Salpeter1955,Hopkins2018}, and a log-normal (or several power-law sections that may be approximated by a log-normal) turnover toward smaller scales, into the brown-dwarf regime \citep{Kroupa2001,Chabrier2005}, which is difficult to constrain exactly, due to the uncertainties involved in observing low-mass (low-luminosity) stars. The peak (or characteristic mass) of the IMF is around \mbox{$0.1$--$0.5\msun$}.

\subsubsection{Physical processes}

The IMF is controlled by a combination of physical processes. Gravity and turbulence play a central role \citep{KlessenHeitschMacLow2000,PadoanNordlund2002,HennebelleChabrier2008,Hopkins2013IMF,NamFederrathKrumholz2021}. However, magnetic fields and radiation \citep{PriceBate2009,Bate2009rad,MathewFederrath2020,MathewFederrath2021} are also key ingredients, as they tend to reduce fragmentation. Moreover, magnetic fields produce powerful jets and outflows from the accretion disk around a newborn star, removing mass from the disk and core, and thereby significantly contributing to setting the final mass of a young star \citep{FederrathEtAl2014,GuszejnovEtAl2020}.

\begin{figure}[h]
\centering
\includegraphics[width=0.95\linewidth]{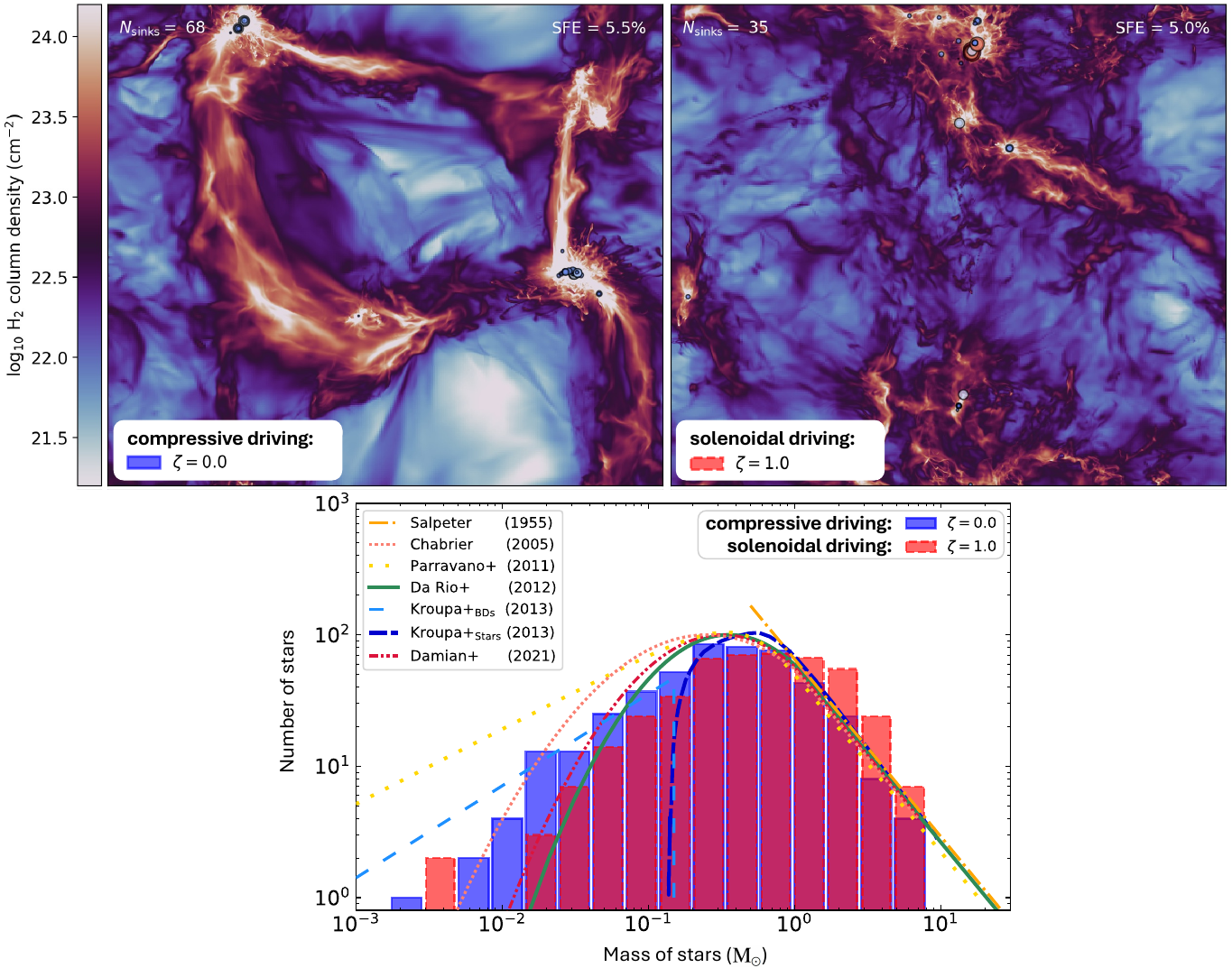}
\caption{Top panels: gas column density in star-formation simulations with purely compressive (curl-free) turbulence driving (left) and purely solenoidal (divergence-free) turbulence driving (right), as defined in Sec.~\ref{sec:pdf}. Young stars are shown as circles, forming in dense gas, primarily at the intersection of filamentary structures \citep{SchneiderEtAl2012}.
Bottom panel: comparison of various observational IMFs together with the IMF obtained in several sets of simulations using the driving modes of turbulence shown in the top panels: compressive driving (blue histogram) and solenoidal driving (red histogram). The curves are the system IMF models based on observational surveys by \citet{Salpeter1955} (dash-dotted), \citet{Chabrier2005} (short-dotted), \citet{ParravanoMcKeeHollenbach2011} (long-dotted), \citet{DaRioEtAl2012} (solid), \citet{KroupaEtAl2013} for brown dwarfs (long-dashed) and stars (short-dashed), and \citet{DamianEtAl2021} (dash-double-dotted). Adapted from figures 2 and 6 of \citet{MathewFederrathSeta2023}.}
\label{fig:imf}
\end{figure}

In \Cref{sec:sfrbasics} we saw that the turbulent density distribution determines the amount of dense gas eligible for star formation, thereby determining the SFR. Similar holds for the IMF, with most modern theories of the IMF relying on the same underlying physics that gives rise to the turbulent gas density distribution described by \Cref{eq:pdf}. Here we highlight one aspect of this distribution, namely, its width, which is crucially determined by the driving mode of the turbulence ($b$ parameter in \Cref{eq:sigs}). Using numerical simulations, \citet{MathewFederrathSeta2023} showed that the IMF depends on the driving mode (solenoidal vs.~compressive) of the turbulence. This is shown in \Cref{fig:imf}. We see that compressive driving produces substantially stronger density fluctuations than solenoidal driving. The IMF resulting from several sets of simulations with different random seeds yields a total of 468 and 445~stars, with a median stellar mass of $(0.4\pm0.1)\msun$ and $(0.6\pm0.2)\msun$ for compressive and solenoidal driving, respectively. This shows that turbulence is a key ingredient for the IMF, and variations in the driving mode of the turbulence may produce significant variations in the IMF.

\subsection{Feedback processes}
While the interplay of turbulence and gravity is a key controller of star formation, as discussed in the previous two subsections, stellar feedback processes also play a crucial role. We broadly distinguish mechanical and radiative forms of feedback.

Mechanical feedback is the redistribution of mass and momentum by jets and outflows from the accretion disk around protostars (see Sec.~\ref{ssec:magnetic_fields}) or from supernova explosions. Jets and outflows are particularly relevant for the SFR and IMF, in that they limit the amount of material that can be accreted onto the protostar by about a factor of 2, therefore slowing down star formation \citep{PadoanNordlund2011,FederrathEtAl2014,Federrath2015}. Moreover, the mechanical nature of this type of feedback can cause coherent accretion streams to break, thereby inducing additional fragmentation, which, together with the direct limiting effect on accretion, leads to an overall reduction of the average stellar mass by a factor of $\sim3$ \citep{FederrathEtAl2014,GuszejnovEtAl2020,MathewFederrath2021}.

Radiative feedback describes the heating, ionization, and/or radiation pressure induced by stars. This form of feedback, in particular direct radiation pressure and reprocessed ionizing radiation from massive stars, also causes a mechanical effect in that the radiation force can push on the dust \citep{MenonEtAl2023}, forming expanding shells around HII regions, sculpting dense structures such as the Pillars of Creation. Evolved stars drive winds throughout most of their lifetime, re-injecting material (in particular metals), momentum and energy into the ISM. While the aforementioned radiative feedback processes are primarily relevant for massive stars, heating feedback is crucial for all young stars, including low-mass stars. Accretion causes a local ($\lesssim0.1\pc$) heating effect around young stars, which limits fragmentation of the surrounding gas, thereby significantly controlling the low-mass end of IMF \citep{OffnerEtAl2009,Bate2009rad,PriceBate2009,FederrathKrumholzHopkins2017,GuszejnovEtAl2018,MathewFederrath2020}.

Finally, all the mechanical feedback types, as well as the radiative ones that cause a mechanical effect can drive turbulence \citep{Elmegreen2009,FederrathEtAl2017iaus}, thereby closing a feedback loop, in which turbulence is responsible for regulating star formation as described in Sec.~\ref{sec:sfrbasics} and~\ref{ssec:IMF} above.

\section{Observational view of the star formation process}
\label{sec:observations}

In this section, we describe the observational constraints on the formation of a single Solar-like stellar system. Due to the embedded nature of protostellar sources, their studies are mostly conducted at infrared and longer wavelengths, as the young protostars are often in the densest part of the cores from which they form, where extinction inhibits observations at shorter wavelengths. The focus of this section is the protostellar stages (i.e., Class~0 and Class~I objects). The evolution of later stages (pre-main sequence stars),  protoplanetary disks, as well as planet formation, is covered in `Protoplanetary disk origins and free-floating exoplanets', `Protoplanetary disk chemistry and structure', and `Planet formation mechanisms'.

\subsection{Protostellar evolutionary path and classifications} 

The protostellar evolution is divided into classes, based on their observed properties. The different observational properties of protostellar evolutionary characterization are summarized in Table~\ref{tab:protostellar_classes}. These empirical classes of evolution were first introduced based on the observations of a near-infrared spectral index between 2 to 20~$\mu$m defined as

\begin{equation}  \label{eq:alfa_ir}
\alpha _{\textrm{IR}}= \frac{d\,\log ({\lambda F_\lambda})}{d\,\log (\lambda)},
\end{equation} 

\noindent for flux $F_\lambda$ at wavelength $\lambda$ \citep{Lada.Wilking1984}. The spectral index changes as the protostar gains mass, disperses the envelope and forms a protostellar disk. The youngest protostars show a redder (positive) spectral index, with increasing brightness towards longer wavelengths, and more evolved sources have a negative spectral index as the stellar spectral energy distribution (SED) approaches that of a main-sequence star.

However, this characteristic does not account for the existence of even younger objects, Class~0 protostars, which are often too cold and visually extinct to emit in the near-IR regime \citep{Andre.WardThompson.ea1993}. Flat-spectrum sources were also later distinguished as a transition between Class~I and Class~II sources \citep{Greene.Wilking.ea1994}.

\begin{table}[t]
\TBL{\caption{Observational characteristics of protostellar classifications}\label{tab:protostellar_classes}}
{\begin{tabular*}{\textwidth}{@{\extracolsep{\fill}}@{}llll@{}}
\toprule
\multicolumn{1}{@{}l}{Class} & 
\multicolumn{1}{l}{$\alpha _{\textrm{IR}}$} &
\multicolumn{1}{l}{$L_{\textrm{submm}}/L_{\textrm{bol}}$} &
\multicolumn{1}{l@{}}{$T_{\textrm{bol}}$ [K]}\\
\colrule
0 & -- & $\geq 0.5\%$ & $\leq 70$\\
I & $\geq 0.3$ & $< 0.5\%$ & $70 - 650$ \\
Flat Spectrum & $0.3 - -0.3$  & -- & --  \\
II & $-0.3 - -1.6$ & -- & $650 - 2800$  \\
III& $\leq -1.6$ & -- & $\geq 2800$ \\
\botrule
\end{tabular*}}{%
\begin{tablenotes}
\footnotetext[]{Class: protostellar class, $\alpha_{\textrm{IR}}$: the gradient of the spectral energy distribution in the infrared, $L_{\textrm{submm}}/L_{\textrm{bol}}$: the ratio of the luminosity in the sub-millimeter and the bolometric luminosity, $T_{\textrm{bol}}$: bolometric temperature.}
\end{tablenotes}
}%
\end{table}

\begin{figure}[b]
\centering
\includegraphics[width=.95\textwidth]{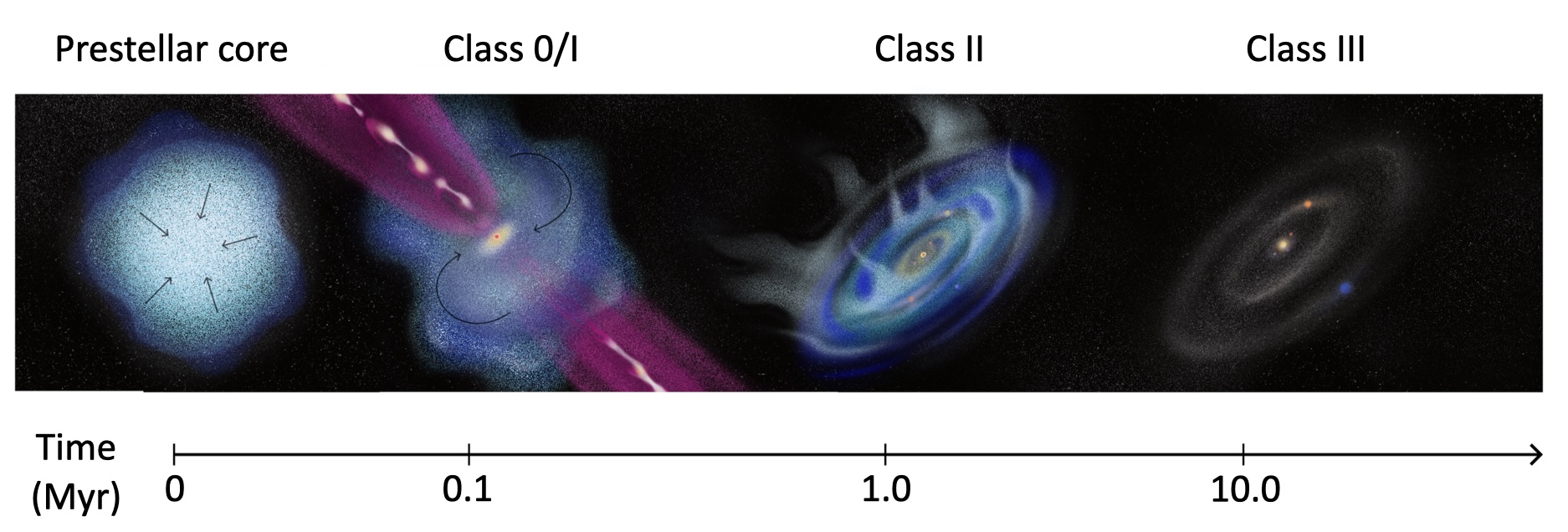}
\caption{An illustrated overview of protostellar evolutionary classes. In the collapse stage, the infall motions create a dense central region in the prestellar core. Class 0/I stage (protostellar stage) is associated with powerful outflows and jets accompanied by the most vigorous accretion; this is also the stage for protoplanetary disk forms. In the Class II stage, the disk is cold and quiescent, associated with disk winds, this is also where embedded planets are detected. In Class III, the disk disperses and a residual dusty debris disk is present.}
\label{fig:physical_protostellar}
\end{figure}

Protostars have been also categorized by their bolometric temperature T$_{\rm bol}$, which is established as the temperature of a blackbody with the same mean wavelength as the SED of a protostar \citep{Myers.Ladd1993}. Based on that classification, the Class~0 stage can be distinguished as having $T_{\rm bol} \leq 70\K$. Both of the described methods, however, rely on observable properties of the system, where, for example, the inclination of the protostellar disk can alter the measured infrared spectral index \citep{Whitney.Wood.ea2003}. Another method involves comparing the contribution of sub-millimeter luminosity to the bolometric luminosity of the source, where most embedded sources have at least 0.5\% of their luminosity above $350\,\mu$m contribution \citep{Andre.WardThompson.ea1993}.

Another way to classify protostellar sources is by their physical parameters instead of observed properties, which provide more descriptive characteristics of the state of the system \citep{Whitney.Wood.ea2003, Robitaille.Whitney.ea2006}. These physical classifications are illustrated in \Cref{fig:physical_protostellar}. In Class~0, most of the system's mass is still in the envelope; Class~I marks the transition where disk mass is comparable to or greater than the mass of the envelope, while most of the system's mass is already in the central star; Class~II sources have a negligible envelope, with the gaseous disk still present, but its mass is much lower than the mass of the central star; by Class~III the star is a pre-main sequence object and the disk is gas-less and of negligible mass. 

\begin{figure}[t]
\centering
\includegraphics[width=.55\textwidth]{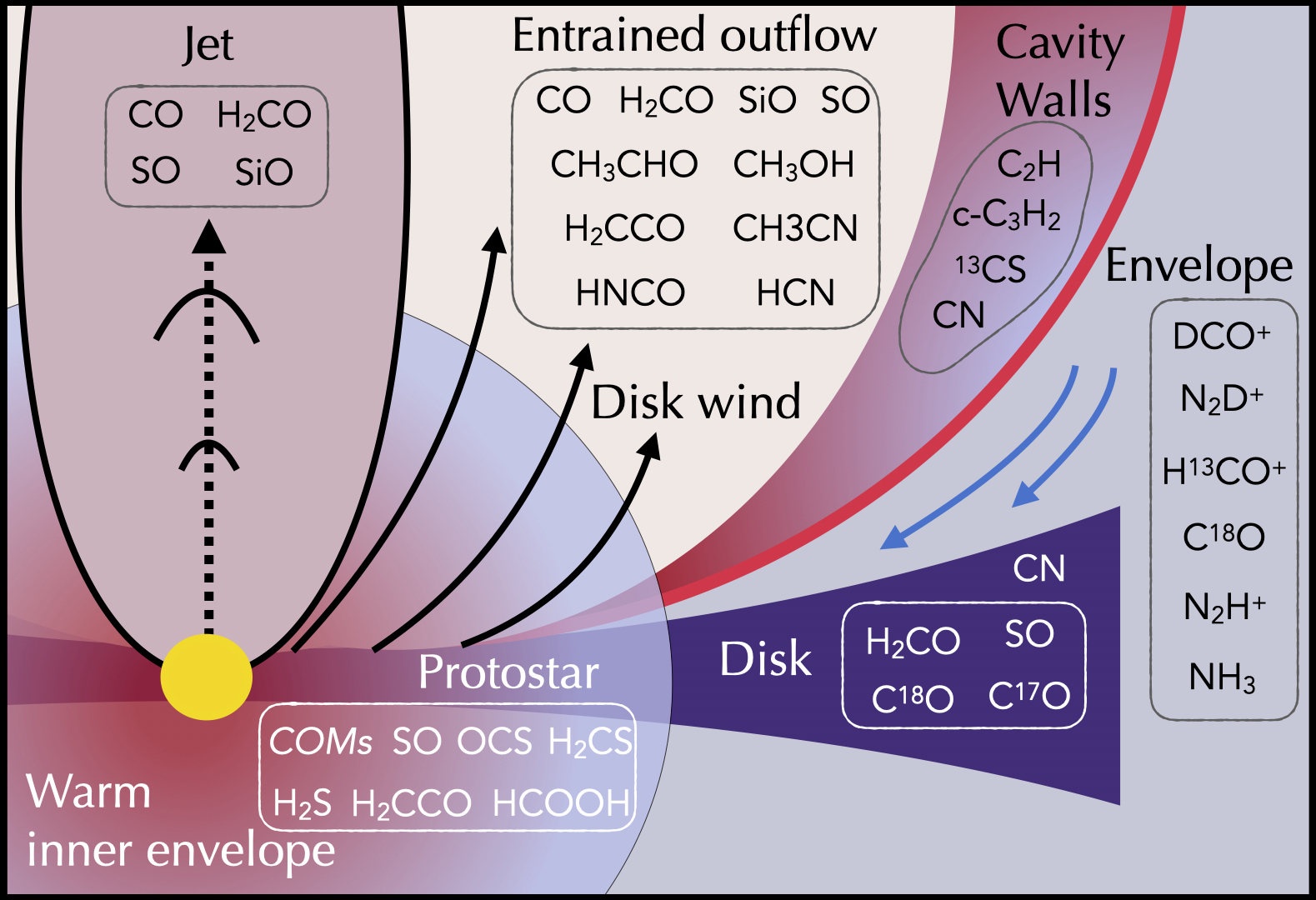}
\caption{Schematic of protostellar physical components with molecules and their associated emission/absorption lines that can be used to probe disk structure, chemistry, and dynamics through sub-millimeter spectroscopy. From \citet{Tychoniec.vanDishoeck.ea2021}, Reproduced with permission from Astronomy \& Astrophysics, \textcopyright ESO.}
\label{fig:tracers}
\end{figure}

A protostellar system comprises different physical components that can be observationally characterized using various molecular tracers. These tracers are summarised in \Cref{fig:tracers}. In the following sections, we describe the key characteristics and evolutionary trends in each of those components.

\subsection{Protostellar envelope}

Sub-millimeter single-dish and interferometric continuum observations, sensitive to cold grains, are widely used tools to recover the dust structure of the envelopes. Observations find that protostellar envelopes have radii of several $1000\au$. Inferred from dust emission, the density profiles of the protostellar envelopes are often found following a radial density profile close to $\rho \propto r^{-2}$ \citep{Looney.Mundy.ea2003,Maury.Andre.ea2019}, which is consistent with theoretical predictions described in \cite{Larson1969}, but steeper profile closer to $\rho \propto r^{-3/2}$ of the inside-out collapse \citep{Shu1977} is also observed \citep{Kristensen.vanDishoeck.ea2012}. Dust properties are typically similar to the interstellar medium; however, in the inner envelope, signatures of grain growth can be observed \citep{Galametz.Maury.ea2019}.

Gas in the protostellar envelopes is traced almost exclusively at (sub-)millimeter wavelengths due to the very low temperatures of the order of $10-20\K$. Velocity-resolved observations of emission lines can trace the infall and rotation of the envelope and can be used to constrain the angular momentum \citep{Gaudel.Maury.ea2020}. At densities of $10^4-10^5\cm^{-3}$, the freeze-out timescales of the gas become shorter than the envelope lifetime, and certain gas species sublimate onto the dust grains. For example, the freeze-out temperature of carbon monoxide (CO) is $20-25\K$, while water (H$_2$O) freezes at temperatures below $100\K$. The depletion of CO and H$_2$O from the gas to the ice phase causes a rise of emission of molecules, which otherwise are efficiently destroyed through reactions in the gas-phase with CO. Such tracers are DCO$^+$ and N$_2$H$^+$, which are tracers of CO freeze-out and H$^{13}$CO$^+$, which tracers H$_2$O freeze-out \citep{Hogerheijde.vanDishoeck.ea1997, Tobin.Hartmann.ea2011}. The frozen molecules cannot be traced with emission spectroscopy, but they absorb the light, especially in the infrared regime. Combined with the laboratory characterization of ice mixtures, a detailed composition of the ice mantle in envelopes around protostars can be obtained \citep[see][for review]{Boogert.Gerakines.ea2015}.

The envelope dissipates during protostellar evolution as the material is delivered to the disk and star. At the same time, protostellar outflows and jets open up a cavity wall and expel a large amount of material from the system. On the other hand, streamers of gas from the larger cloud scales can replenish the envelope with material at various stages of evolution \citep[e.g.,][]{Pineda.SeguraCox.ea2020}.

\subsection{Outflows and jets}

Outflows are one of the first signs of the new star being born as they expel gas away from the deeply embedded protostar. As the outflow propagates at supersonic velocities, it creates shocks with the surrounding medium. Therefore, high-temperature traces such as H$_2$ rotational transitions are commonly used to study shocked gas. Shocks disrupt dust mantles and cores, releasing material that would, in quiescent ISM conditions, remain in the solid phase. Therefore, SiO molecular gas or atomic and ionized emission from Si, Fe, and Ni is observed in shocked gas.

Observationally, outflows are typically divided into the high velocity ($>30\kms$), highly collimated component often called jets, and the low velocity ($<30\kms$) wind angle component sometimes called winds. The low-velocity component is expected to trace the envelope material entrained by the faster component, or the disk wind which is the gas directly released from the protoplanetary disk. The low-velocity outflow is traced by rotational transitions of CO. In some young outflows, the outflow can also be traced by more complex species such as CH$_3$OH and H$_2$CO, which trace the sputtering of grains in low-velocity shocks at the outflow cavity walls. 

Detailed studies of jet kinematics can inform about their precise physical origin and mechanism (see \ref{ssec:magnetic_fields}. The chemical content of the jets undergoes evolution. Molecules such as CO, SiO, and SO are mostly detected in very young Class~0 sources. This is likely because high number densities of the order of $10^6\cm^{-3}$ are required for efficient gas-phase formation of molecules from initially atomic material. Further into the evolution, the neutral ([O~I], [Ni~I], [Cl~I], [S~I]) and ionized ([Fe~II], [Ne~II], [Ar~II]) components of the jet become dominant \citep{Nisini.Santangelo.ea2015}. Prominent refractory contents of the jet material suggest that jets either launch from the inner regions of the disk or that dust grains are launched and efficiently destroyed in the jet. 

Apart from chemical evolution, jets and outflows also significantly change their energetic and mass output during protostellar life. Young outflows are characterized by the most energetic outflows, and the total outflow force is found to be correlated with protostellar luminosity, indicating a strong relation between accretion and ejection activity of the protostar \citep{Bontemps.Andre.ea1996}. This correlation between outflow and accretion rate is used to design simulation sub-resolution models of jets and outflows \citep{CunninghamEtAl2011,FederrathEtAl2014,GuszejnovEtAl2020}.

Since the outflows are expected to remove angular momentum, observations of the rotational signature is one of the crucial observations. Rotation of the jet and wind has been observed \citep{Bjerkeli.vanderWiel.ea2016, Lee.Ho.ea2017}, indicating that the angular momentum is indeed removed with the outflow.

Jets launched from the inner regions of the protoplanetary disks often form internal shocks, which are characterized by high densities, where molecules can efficiently form. Those shocks are caused by internal variations of the jet velocity, which occur due to accretion variability. Because of that, jets are fossil records of the accretion process, revealing that the protostellar accretion process is highly variable in nature \citep{Lee2020}.

\subsection{Protostellar accretion}
\label{ssec:accretion}

Most of the stellar mass is assembled during the early stages of evolution (Class 0/I). Direct observations of the protostar remain a challenge for observations since the protostars are deeply embedded. Nevertheless, in recent years significant progress has been made to extract stellar properties \citep{Fiorellino.Tychoniec.ea2023}. Hydrogen recombination lines, which are tracers of high-density and high-temperature gas, are used to probe the accretion onto the protostar. With a combination of bolometric luminosity estimates and infrared photometry, stellar properties can be constrained. 

Measured accretion rates are often lower than expected, considering the duration of accretion and the final masses of stars from the initial mass function \citep{Kenyon.Hartmann.ea1990}. This discrepancy between the observed accretion rate of young stars being significantly lower than expected from models is called the Luminosity Problem. A solution to this problem is that protostars accrete a significant portion of their mass during periods of high accretion, such as outbursts or in the initial stages of protostar formation. Protostellar accretion is, therefore, a highly variable process that evolves dramatically during protostellar life \citep{Fischer.Hillenbrand.ea2023}.

\subsection{Embedded disk}

In the inner regions of the envelope, the velocity profile changes as the forming circumstellar accretion disk follows Keplerian rotation. Several young disks have their Keplerian rotation characterized in observations. However, it remains a challenge as most of the dust disks are small, with radii $\leq50\au$ \citep{Maury.Andre.ea2019}. With different tracers such as formaldehyde (H$_2$CO) or optically-thin isotopologues of CO, it is possible to study the temperature of the disk \citep{vantHoff.Harsono.ea2020}.

Dust masses of the young Class~0 and Class~I disks are fundamental to estimating the total budget of building blocks of planets. However, they are difficult to constrain as the young disks are optically thick and hard to discern from the surrounding envelope. Observations at longer wavelengths $\sim$1$\cm$ can mitigate those issues and have been used to constrain masses of the order of 50 to 150~Earth masses \citep{Tychoniec.Manara.ea2020}.
This is a factor of 5 to 20 more than typical masses of Class~II disks \citep{Ansdell.Williams.ea2017}.

The available mass budget, grain growth observed in Class~I systems, and substructures omnipresent in Class~II disks suggest that planet formation should already begin early. These structures, such as gaps and spirals, are rarely observed in Class~0, while they appear to be more common in the Class~I stage, suggesting an evolution of disks potentially shaped by planets \citep{Ohashi.Tobin.ea2023}. For further details on protoplanetary disks and planet formation, we refer to `Protoplanetary disk chemistry and structure' and `Planet formation mechanisms'.

\section{Multiplicity and the formation of binary/multiple star formation}
\label{sec:multiplcity}

\begin{figure}[t]
\centering
\includegraphics[width=0.8\textwidth]{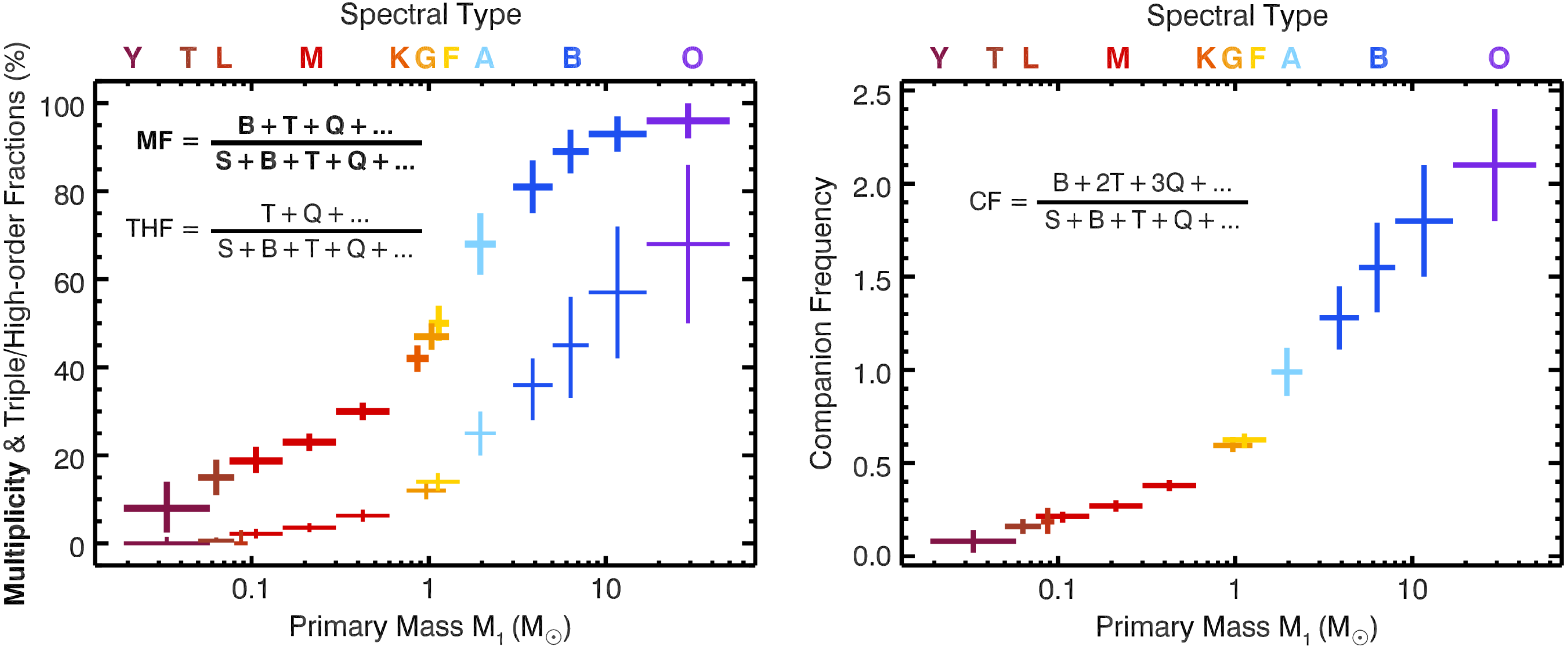}
\caption{\emph{Left}: observed multiplicity fraction as a function of star mass. The fraction of stars in binary or higher-order, or triple or higher-order systems is shown by the thick and thin crosses, respectively. \emph{Right}: companion frequency, or the average number of stellar companions a star has as a function of star mass. Reproduced with permission from \citet{offner_origin_2023}.}
\label{fig:multiplcity_stats}
\end{figure}

\Cref{sec:physical_processes} and \ref{sec:observations} have focused on the formation of a single star, however many stars exist in binary or multiple star systems (\citet{offner_origin_2023}; See also `Observing binary stars'). \Cref{fig:multiplcity_stats} compiles observational surveys of main sequence stars, with the left panel showing the fraction of stars of mass $M$ that are in binary or higher (thick crosses), or triple or higher (thin crosses) star systems. The right panel shows the companion frequency as a function of star mass. We see that many stars can exist in binary or multiple-star systems, with more massive stars being more likely to have companions. The actual fraction of all stars that are in multiple star systems is sensitive to the initial mass function (see \Cref{ssec:IMF}), however, it is accepted that a significant number of stars are in multiple star systems, and their formation must not be ignored when understanding star and planet formation.

We also find that most stars are born with a companion, with multiplicity being highest in the protostellar Class~0 (see \Cref{tab:protostellar_classes}), decreasing as we look at more evolved protostars. This means that many of the stars in these Class~0 multiple-star systems will interact and get ejected as they evolve towards the main sequence, or maybe even merge to form more massive stars \citep{bally_birth_2005}. These interactions early on can affect the disks around the protostars and affect the sites of planet formation. Many stars that are single on the main sequence may have begun their life in a multiple-star system and were ejected through complex orbital dynamics.

\begin{figure}[t]
\centering
\includegraphics[width=0.45\textwidth]{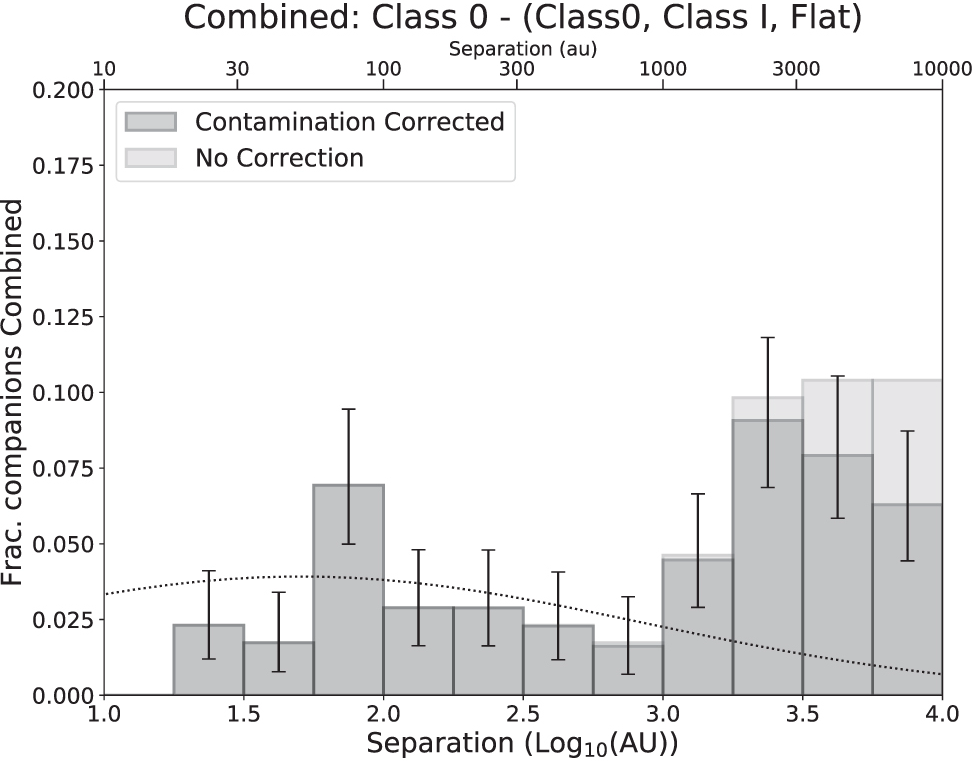}
\caption{Observed separation distribution of young multiple star systems, with at least one Class~0 object. This combines observations from the Perseus and Orion star-forming regions. The thin dotted line is the separation distribution of solar-type field stars \citep{raghavan_survey_2010}. Reproduced with permission from \citet{tobin_vlaalma_2022}.}
\label{fig:protostellar_multiplicity}
\end{figure}

Observations of separations in young binary and multiple star systems in star-forming regions find a bimodal distribution with one peak at $\sim$100$\au$ and another at $\sim$3000$\au$, as seen in \Cref{fig:protostellar_multiplicity}. When this bimodal distribution was first observed, the origin of the two peaks was attributed to two formation pathways for multiple star systems: 1.~pre-stellar core fragmentation, and 2.~circumstellar disk fragmentation.

\subsection{Core fragmentation}

\begin{figure}[t]
  \centering
  \begin{subfigure}{.55\linewidth}
    \centering
    \includegraphics[width=\linewidth]{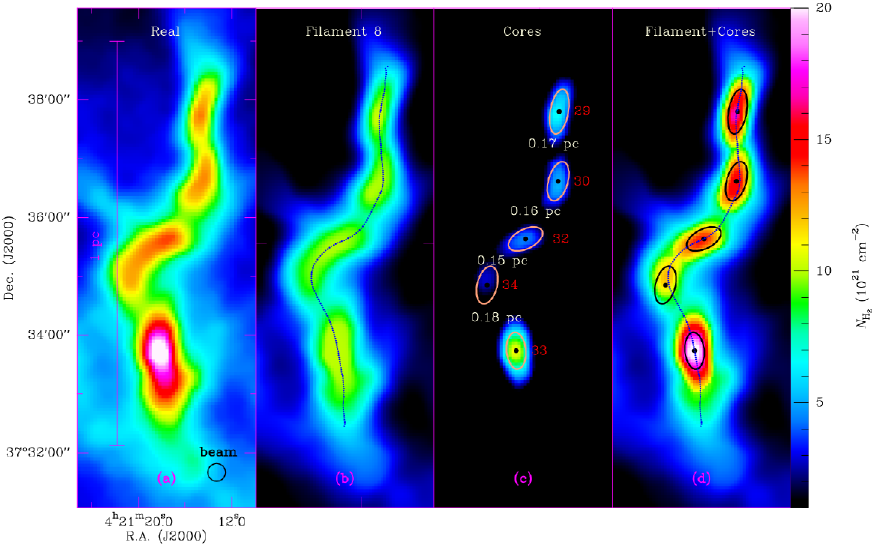}
    \caption{Filament fragmentation in the California molecular cloud. (a) the true observations, (b) the extracted filament, (c) the identified pre-stellar cores, and (d) the reconstructed observations with the modeled filament and cores. From \citep{zhang_fragmentation_2020}, reproduced with permission from Astronomy \& Astrophysics, \textcopyright ESO.}
    \label{subfig:filament_frag}
  \end{subfigure}%
  \hspace{1em}
  \begin{subfigure}{.43\linewidth}
    \centering
    \includegraphics[width=\linewidth]{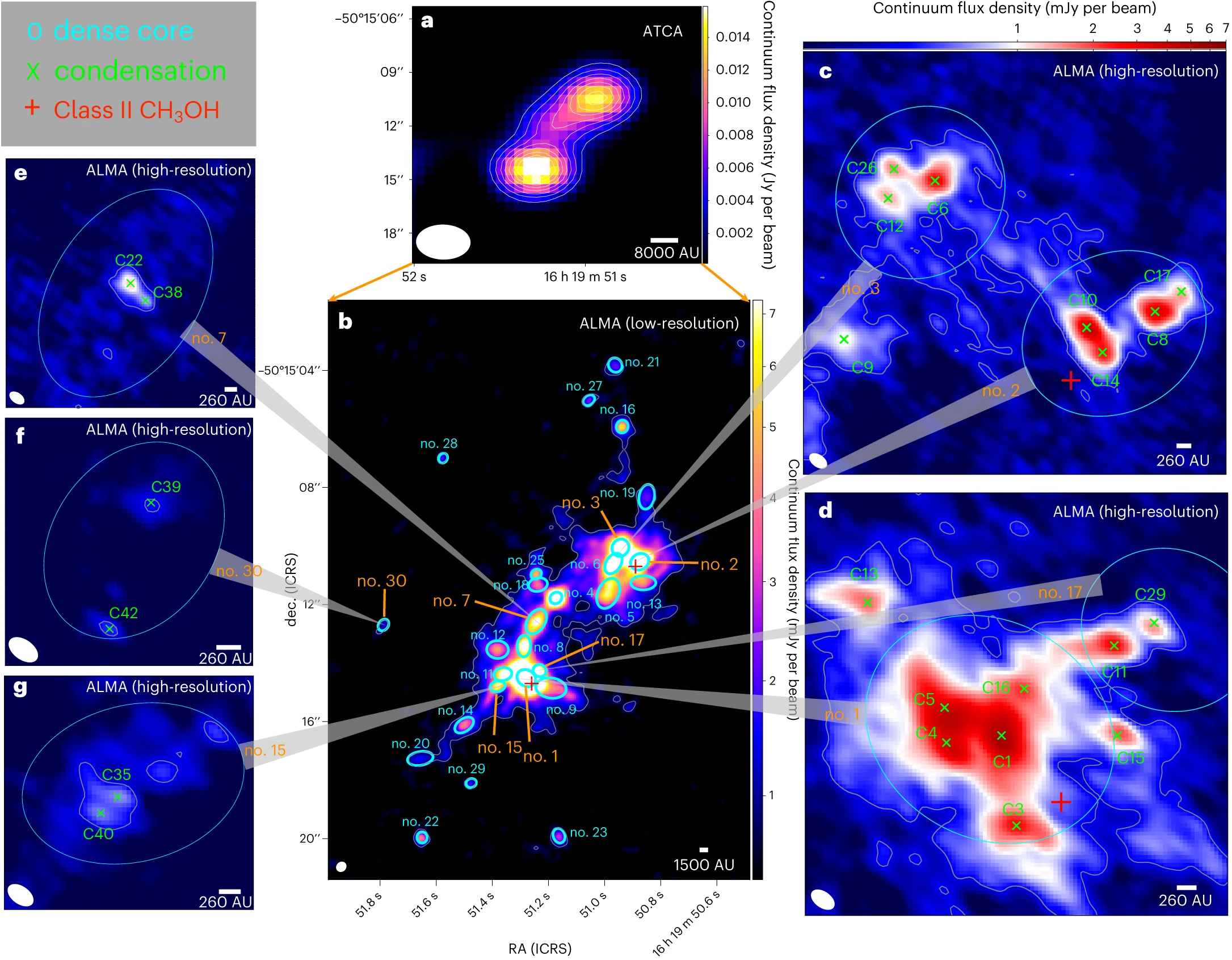}
    \caption{Hub fragmentation in G333. Reproduced from \citep{li_observations_2024}. (a) shows the low-resolution ATCA observations, (b) shows the low-resolution ALMA observations revealing further fragmentation, and the remaining panels show high-resolution ALMA observations of parts of the hub, showing further hierarchical fragmentation.}
    \label{subfig:hub_frag}
  \end{subfigure}%

  \caption{Observations of fragmentation in different environments.}
  \label{fig:core_frag_obs}
\end{figure}

Core fragmentation was used to explain the separation peak at $\sim$3000$\au$ because this formation pathway acts on larger scales of 100s to 1000s of $\au$. As stated throughout this chapter, molecular clouds are turbulent, and this turbulence can create over-densities that make pre-stellar cores collapse to form a star. However, pre-stellar cores also have sub-sonic turbulence, which may seed further over-densities that can fragment to form stars. The description of pre-stellar collapse in \Cref{sec:physical_processes} starts with a spherical cloud, however, turbulence in the ISM can seed the formation of filaments \citep{Federrath2016}, from which most pre-stellar cores fragment. This is seen in \Cref{subfig:filament_frag}, with an observed filament (leftmost panel) versus a modeled filament and cores (rightmost panel). This elongation adds asymmetry, which can seed fragmentation, along with the turbulent nature of the cores.

Fragmentation in hubs, where filaments intersect is also observed. This is seen in \Cref{subfig:hub_frag} where at low resolution (center panel) fragmentation is observed, and at higher resolution, further hierarchical fragmentation is also observed. The filaments that feed these hubs inject turbulent energy, which can lead to fragmentation.

The fragments that form along a filament can dynamically fall towards each other because the relative velocity of the fragments to each other is low, and stars that form from core fragmentation in hubs will likely also experience complex dynamical interactions. While the larger peak in the separation distribution at $\sim$3000$\au$ is attributed to core fragmentation, many multiple star systems that form via this pathway often inspiral to smaller separations, even down to $<100\au$ \citep{kuruwita_contribution_2023}.

\subsection{Disk fragmentation}

As described in \Cref{ssec:magnetic_fields}, circumstellar disks are a natural consequence of pre-stellar core collapse, and under the right conditions, fragmentation can occur within these disks to form new stars. The separation peak at $\sim$100$\au$ has been attributed to disk fragmentation because this formation pathway acts on disk scales. Circumstellar disks can extend up to $\sim600\au$, with mean disk sizes in the Class~0/I stage being around $\sim75\au$ \citep{tsukamoto_role_2022}.

The Toomre~$Q$ \citep{toomre_gravitational_1964} is a quantity that is often used to measure the stability of a disk, defined as

\begin{equation}
    Q = \frac{\cs\Omega}{\pi G \Sigma},
    \label{eqn:Toomre_Q}
\end{equation}

\noindent where $\cs$ is the sound speed, $\Omega$ is the angular frequency, and $\Sigma$ is the gas surface density of the disk. A parcel of gas is considered to be stable if $Q\gg1$, and unstable and prone to collapse if $Q\ll1$. The Toomre~$Q$ essentially measures a ratio of how pressure and rotationally supported the disk is against its own gravity. A disk that is Toomre unstable can become stable by either rotating faster (increase $\Omega$), reducing surface density, or being hotter; i.e., higher $\cs$. This definition ignored other forms of support against fragmentation, specifically any type of non-thermal pressure. Sources of non-thermal pressure include magnetic pressure, turbulent pressure, and, radiation pressure. The latter two in particular, as well as magnetic tension, however, can have significant anisotropic effects, so simply adding them as an isotropic pressure contribution to $Q$ may be too simplistic. A Toomre~$Q$ that has magnetic pressure added has been derived from MHD simulations \citep{forgan_fragmentation_2017}, and is defined by multiplying the Toomre~$Q$ by a scaling factor that arises from adding the thermal and magnetic pressures \citep[see eqs.~17 and~18 in][]{FederrathKlessen2012},

\begin{equation}
    Q_B = Q\sqrt{1 + \beta^{-1}},
    \label{eqn:Mag_Toomre_Q}
\end{equation}

\noindent where $\beta$ is plasma beta, as defined in \Cref{sec:pdf}.

The ideal disks for fragmentation are massive, cold disks, which are relatively rare. Radiation feedback from the central star, is expected to provide thermal support against disk fragmentation \citep{offner_effects_2011}. However, stars accrete episodically (see \Cref{ssec:accretion}), therefore circumstellar disks can go through cycles of heating and cooling, and if the time between accretion events is sufficiently long, disks may cool temporarily to become unstable \citep{stamatellos_episodic_2012}.

\subsection{Evolution of young multiple star systems}

After fragmentation into binary or multiple-star systems, these stars also interact. Simulations of the formation of eccentric binaries find that accretion bursts can be triggered at periastron (the closest separation) because the companion star disrupts the circumstellar disk \citep{kuruwita_dependence_2020}. Accretion bursts can also be triggered by the flyby of unbound stars \citep{borchert_rise_2022}, which is not uncommon in clustered star-forming environments. Most binaries have orbital periods that are significantly longer than a human lifetime, therefore, it has been difficult to observe directly companion-triggered accretion, but there are a handful of short-period young binaries where companion-triggered accretion is observed over multiple orbits \citep{mathieu_classical_1997}.

Interactions between stars can also truncate circumstellar disks. Simulations find that the radius of circumstellar disks is truncated to approximately a third of the binary separation \citep{artymowicz_dynamics_1994}. Thus, truncation can shorten the lifetime of the disk and potentially hinder planet formation.

While circumstellar disks (the disks around individual stars) can be truncated or destroyed by binary-star interactions, simulations find that the formation of circumbinary disks is ubiquitous. Circumbinary disks can form either via the inspiral of binaries formed through core-fragmentation \citep{kuruwita_role_2019}, or through disk fragmentation \citep{tokovinin_architecture_2021}. Observations find that many of the largest protostellar disks are circumbinary disks \citep{harris_resolved_2012}, and some are unusually old ($>10\Myr$), for example, AK~Sco ($18\pm1\Myr$; \citealt{czekala_disk-based_2015}), HD~98800~B ($10\pm5\Myr$; \citealt{furlan_hd_2007}) and V4046~Sgr ($12–23\Myr$; \citealt{rapson_combined_2015}). The size and persistence of circumbinary disks may provide an ideal environment for planet formation. For details on accretion from circumbinary disks, we refer the reader to `Circumbinary Disk Accretion'.

Young multiple-star systems can experience complex orbital dynamics such as higher-order systems ejecting companions, and new multiple-star systems forming through dynamical capture. Approximately one-third of binaries are estimated to not have been born together based on observations \citep{murillo_siblings_2016} and simulations \citep{kuruwita_contribution_2023}. Dynamical capture is likely to be easier in star-forming environments because these young stars are actively accreting from their gaseous environments, which produce dynamical drag. Simulations of binaries that formed via core fragmentation also find that in-spiraling halts when the binary is no longer embedded in a dense gaseous environment \citep{kuruwita_contribution_2023}, highlighting that early stellar dynamics are strongly influenced by gas dynamics. Once these young multiple-star systems have accreted mass and are no longer embedded, they are not expected to evolve much dynamically. For details on the evolution of multiple are systems after their initial formation, we refer readers to `Evolution of binary stars'.

\section{Conclusions}

Stars form in turbulent environments with a complex interplay of different physics. At the beginning of this chapter, we reviewed the role of gravity, hydrodynamics, radiation, and magnetism in the collapse of a pre-stellar core into a star. We find that the collapse of a core by gravity is counteracted by gas pressure, radiation feedback, and magnetic pressure on different scales, but efficient radiative cooling, and angular momentum removal by magnetic fields, jets, and outflows can also aid pre-stellar collapse. We provided a summary of the physics of molecular clouds in which stars form, and how the interplay of gravity, turbulence, magnetic fields, and stellar feedback in the form of jets/outflows and radiation controls the star formation rate (SFR) and the initial mass function (IMF) of stars. We then reviewed what observations can tell us about the star formation processes. Sub-millimeter and infrared studies reveal a complex interplay of different components of the protostellar systems. Observations of gas kinematics can constrain theoretical predictions on the origin of protostellar jets and outflows, while thermal dust continuum observations deliver constraints on the onset of planet formation. Finally, we highlighted how the formation of multiple star systems complicates our single-star picture of star formation. We emphasize that most stars are born with companions and why and how this may affect planet formation.

\begin{ack}[Acknowledgments]{}
R.L.K.~acknowledges funding from the Klaus Tschira Foundation.
C.F.~acknowledges funding provided by the Australian Research Council (Discovery Project DP230102280), and the Australia-Germany Joint Research Cooperation Scheme (UA-DAAD). L.T. is supported by the Netherlands Research School for Astronomy (NOVA).
\end{ack}

\newcommand{\physrep}{Physics Reports}
\newcommand{\ssr}{Space Science Reviews}
\newcommand{\apjl}{The Astrophysical Journal Letters}
\newcommand{\apjs}{The Astrophysical Journal Supplement Series}
\newcommand{\apj}{The Astrophysical Journal}
\newcommand{\aj}{The Astronomical Journal}
\newcommand{\aap}{Astronomy \& Astrophysics}
\newcommand{\mnras}{Monthly Notices of the Royal Astronomical Society}
\newcommand{\araa}{Annual Review of Astronomy \& Astrophysics}
\newcommand{\pasp}{Publications of the Astronomical Society of the Pacific}
\newcommand{\pasa}{Publications of the Astronomical Society of Australia}
\newcommand{\pre}{Physical Review Letters}
\newcommand{\prd}{Physical Review D}
\newcommand{\nat}{Nature}
\newcommand\aapr{Astronomy and Astrophysics Reviews}
\newcommand{\jcp}{Journal of Computational Physics}
\newcommand{\jfm}{Journal of Fluid Mechanics}
\newcommand{\rmp}{Reviews of Modern Physics}
\newcommand{\prl}{Physical Review Letters}
\newcommand{\na}{New Astronomy}
\newcommand{\jqsrt}{Journal of Quantitative Spectroscopy and Radiative Transfer}

\bibliographystyle{Harvard}
\bibliography{kuruwita,federrath,tychoniec}

\end{document}